\documentclass[twocolumn,showpacs,preprintnumbers,%
               aps,prd]{revtex4-1}
\usepackage[dvips]{graphicx}
\usepackage{amsmath,amssymb}
\usepackage{color}

\newcommand{\rmd}{\mathrm{d}}
\newcommand{\rme}{\mathrm{e}}
\newcommand{\rmi}{\mathrm{i}}

\newcommand{\bp}{\boldsymbol{p}}
\newcommand{\bq}{\boldsymbol{q}}
\newcommand{\bgamma}{\boldsymbol{\gamma}}

\newcommand{\diag}{\text{diag}}
\newcommand{\bmu}{\bar{\mu}}
\newcommand{\bnu}{\bar{\nu}}

\newcommand{\calM}{\mathcal{M}}

\begin{document}

\title{Magnetic-field Induced Screening Effect
       and Collective Excitations}

\author{Kenji Fukushima}
\affiliation{Department of Physics, Keio University,
             Kanagawa 223-8522, Japan}

\begin{abstract}
  We explicitly construct the fermion propagator in a magnetic field
  background $B$ to take the lowest Landau-level approximation.  We
  analyze the energy and momentum dependence in the polarization
  tensor and discuss the collective excitations.  We find there appear
  two branches of collective modes in one of two transverse gauge
  particles;  one represents a massive and attenuated gauge particle
  and the other behaves similar to the zero sound at finite density.
\end{abstract}
\maketitle


Systems in the extreme environment with intense fields
\cite{Dunne:2004nc} are attracting theoretical interest recently
because experimental developments are making it a real possibility
to observe exotic phenomena in strong fields.  In quantum
electrodynamics (QED) a pair production is expected under a strong
electric field $E$, which is called the Schwinger mechanism, and the
critical strength of $E$ necessary for the electron-positron
production is given by an order of $|eE|\sim m_e^2$, i.e.\
$E\sim m_e^2/e\sim 10^{16}V/\text{cm}$.  Such a strong field is still
beyond the experimentally reachable limit.  It is known that
optimization of spatial and temporal profiles of the pulse laser could
significantly lower the critical $E$, which enhances a chance further
to realize the Schwinger pair production in the laboratory
\cite{Schutzhold:2008pz}.

A magnetic field $B$ is also of paramount importance.  Of course a
time-dependent $E$ would be generally accompanied by $B$.  Moreover,
the effect caused by $B$ itself is interesting enough;  one well-known
example is a phenomenon called the magnetic catalysis
\cite{Gusynin:1994re} which favors a $B$-induced chiral condensate.
The birefringence in a medium of strong $B$ can also be a sizable
effect for $|eB|\sim m_e^2$, which requires
$B\sim m_e^2/e\sim 10^{13}\;\text{gauss}$.  This is again far beyond
the limit of experimentally obtainable intensity, while this critical
$B$ could be of order of the surface strength on the neutron star or
magnetar in nature.

One might have thought that it is difficult to attain such large $B$
that any new phenomenon can be realized in the laboratory.  A new
possibility was then brought from physics of the relativistic
heavy-ion collision experiment which aims to create a new state of
matter out of quarks and gluons (called the quark-gluon plasma).  The
heavy-ion collision experiments have turned out to be successful
particularly at RHIC in BNL and more data is coming from LHC in
CERN.\ \  Because heavy ions are positively charged and move at almost
the speed of light, they can generate a gigantic magnetic field if
their collision geometry is non-central.  It was shown that the
maximum $B$ reached at RHIC energy is as large as
$|eB|\sim m_\pi^2\sim 10^{18}\;\text{gauss}$
\cite{Kharzeev:2007jp,Skokov:2009qp}. This is no longer a QED energy
scale but should be already relevant to physics of quantum
chromodynamics (QCD) such as the chiral magnetic effect, the chiral
vortical effect, the chiral separation effect, and so on
\cite{Kharzeev:2007jp,Metlitski:2005pr}.

Physical properties of QED and QCD matter in an intense $B$ field are
under active investigations by means of analytical and numerical
studies.  In the Monte-Carlo simulation an introduction of strong $B$
is feasible and various observables have been measured in the
lattice-QCD calculation \cite{Buividovich:2009wi} as well as model
studies on the QCD phase transitions \cite{Mizher:2010zb} and on the
color superconducting state of dense quark matter
\cite{Ferrer:2005vd}.  In this work we will explore the $B$-induced
screening effect in an analytical approach, which is analogous to the
screening in the finite-temperature field theory.
\vspace{3mm}


First of all let us write down the fermion propagator.  This part
might look technical but explicit calculations are useful to clarify
how the momentum conservation holds.  Using Ritus' method
\cite{Ritus:1972ky} we can explicitly solve the free Dirac equation.
In the following we use Landau's gauge fixing, i.e.\ $A^0=A^1=A^3=0$
and $A^2=Bx$, which describes $B$ in the third direction.  We here
assume $eB>0$, though we can easily relax this with minor
modification.  We define the projection matrix \cite{Ferrer:2005vd} as
$P_k(x) = \diag\bigl[f_{k+}(x), f_{k-}(x), f_{k+}(x), f_{k-}(x)\bigr]$
with the Landau wave-functions;
\begin{equation}
 \begin{split}
 f_{k+}(x) &= \phi_k(x^1-p^2/eB) \quad (k=0,1,\dots) \\
 f_{k-}(x) &= \phi_{k-1}(x^1-p^2/eB) \quad (k=1,2,\dots) .
 \end{split}
\end{equation}
These are the expressions for $eB>0$ and $f_{k+}$ and $f_{k-}$ should
be swapped if $eB<0$.  Here $\phi_k(x)$ is the wave-function for
the harmonic oscillator with $k$ quanta.

One can then prove that the background of $A^2=Bx$ is eliminated with
this projection matrix and the momentum is replaced accordingly as
$\bp \to \tilde{\bp} = (0,-\sqrt{2eB k},p^3)$.  Now that we have the
complete basis of the solutions of the Dirac equation it is
straightforward to write down the two-point functions in a standard
procedure;
\begin{align}
 & \langle\psi(x)\bar{\psi}(y)\rangle \notag\\
 & = \int\frac{\rmd p^2\,\rmd p^3}{(2\pi)^2}\sum_k
  \frac{\rme^{-\rmi \omega_p(x^0-y^0)+\rmi p^2(x^2-y^2)
   +\rmi p^3(x^3-y^3)}}{2\omega_p} \notag\\
 &\qquad\times P_k(x)(\omega_p\gamma^0 -\tilde{\bp}\cdot\bgamma+m)
  P_k(y) ,
\end{align}
which represents the particle propagation from $y$ to $x$.  We note
that $\omega_p=\sqrt{\tilde{\bp}^2+m^2}$ is the particle energy.  In
the same way we get $\langle\bar{\psi}(y)\psi(x)\rangle$ for the
anti-particle propagation from $x$ to $y$, from which we can get the
retarded, advanced, and Feynman propagators.

We cannot, however, express those propagators in a compact form using
the $\rmi \epsilon$-prescription because $P_k$ is not commutative with
$\gamma^2$.  In the lowest Landau-level (LLL) approximation that we
adopt hereafter, the term involving $\gamma^2$ is dropped and the
Feynman propagator for example simplifies into
\begin{align}
 & S^F_0(x,y) = \langle T \psi(x)\bar{\psi}(y)\rangle_0 \notag\\
 & = \int\frac{\rmd p^0 \rmd p^2 \rmd p^3}{(2\pi)^3}
  \rme^{-\rmi p^0(x^0-y^0)+\rmi p^2(x^2-y^2)+\rmi p^3(x^3-y^3)} \notag\\
 &\times \sqrt{\frac{eB}{\pi}}
  \rme^{-\frac{1}{2}eB[(x^1-\frac{p^2}{eB})^2+(y^1-\frac{p^2}{eB})^2]}
  \frac{\rmi P_0}{p_{\bmu}\gamma^{\bmu}-m+\rmi\epsilon} .
\end{align}
with the spin projection matrix; $P_0 = \text{diag}(1,0,1,0)$.  The
Landau zero-mode exists for one spin state only and $P_0$ is nothing
but the projection matrix onto this allowed spin state.  In the above
the index $\bmu$ refers to only the 0th and 3rd components, namely,
$\bar{p}\cdot\gamma=p_{\bmu}\gamma^{\bmu}=p^0\gamma^0-p^3\gamma^3$.
One might have thought that $p^2/eB$ on the exponential could be
dropped in the strong $B$ limit.  We must keep this term, however,
because the integration variable $p^2$ could take any large number.
In fact, as explained below, the momentum conservation is guaranteed
by the $p^2$-integration with this $p^2/eB$ term left.
\vspace{3mm}



\begin{figure}
 \includegraphics[width=.8\columnwidth]{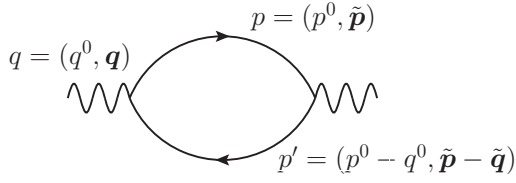}
 \caption{One-loop diagram of the polarization tensor.  The external
   momentum $q$ is not Landau quantized, while the momenta in the loop
   are quantized and replaced by $\tilde{\bp}$ and $\tilde{\bq}$,
   which are further reduced to $\bar{\bp}$ and $\bar{\bq}$ in the
   LLL approximation.}
 \label{fig:diagram}
\end{figure}


We are now ready to go into concrete calculation of the polarization
tensor \cite{Danielsson:1995rh}, which has been discussed repeatedly
in various contexts such as the magnetic catalysis
\cite{Gusynin:1995nb}, the chiral magnetic effect
\cite{Kharzeev:2009pj}, and so on.  In this sense the computation and
the result of the polarization tensor have been already known.  Our
emphasis in this work is thus put in the physical interpretaion
deduced from the energy and momentum dependence in the polarization
tensor.

The one-loop contribution is, from the diagram in
Fig.~\ref{fig:diagram}, given as
$\rmi\Pi^{\mu\nu}(x,y) = -(\rmi e)^2 \bigl[ \gamma^\mu
S_0(x,y)\gamma^\nu S_0(y,x) \bigr]$.  Here we consider the
polarization for the photon, and if it is for the gluon the coupling
constant should be $g$ and the trace in color and flavor space is
necessary.  We shall work in momentum space to calculate,
\begin{equation}
 \rmi\Pi^{\mu\nu}(k,q) = \int\rmd^4 x\, \rmd^4 y\;
  \rme^{\rmi q\cdot x + \rmi k\cdot y}\, \rmi\Pi^{\mu\nu}(x,y) .
\end{equation}
Here $q$ and $k$ are incoming momenta from the external legs attached
to $x$ and $y$, respectively.  Usually, without external magnetic
field, $\Pi^{\mu\nu}(x,y)$ is a function of $x-y$ due to the
translational invariance, which can be Fourier decomposed into
$\exp[-\rmi p\cdot (x-y)]$ and $\exp[-\rmi p'\cdot (y-x)]$ with loop
momenta $p$ and $p'$ (see 
Fig.~\ref{fig:diagram}), so that $x$ and $y$ integrations yield
$(2\pi)^4 \delta^{(4)}(q-p+p')$ and $(2\pi)^4\delta^{(4)}(k-p'+p)$.
Then $p'$ is fixed as $p-q$ from the former constraint and the latter
becomes $(2\pi)^4\delta^{(4)}(k+q)$, meaning that the outgoing
momentum $-k$ is exactly balanced with the incoming momentum $q$.

In the presence of the magnetic field, however, the realization of the
momentum conservation is non-trivial because the translational
invariance is apparently broken (not broken in fact) by the
inhomogeneous vector potential $A^2=Bx$ and $\Pi^{\mu\nu}(x,y)$ is no
longer a function of $x-y$ alone.  Although the final result is known
(see Eq.~(117) in Ref.~\cite{Gusynin:1995nb}), it would be instructive
to make an explicit confirmation in the following.

For $B\neq 0$ the integrations with respect to $x^0$, $y^0$, $x^2$,
$y^2$, $x^3$, $y^3$ are just the standard procedures leading to the
delta-function constraints which lead to $p^{\prime 0}=p^0-q^0$,
$p^{\prime 2}=p^2-q^2$, $p^{\prime 3}=p^3-q^3$, and $k^0+q^0=0$,
$k^2+q^2=0$, $k^3+q^3=0$.  We can further carry out the $x^1$ and
$y^1$ integrations to have
\begin{align}
 & \int\rmd x^1\,\sqrt{\frac{eB}{\pi}}
  \rme^{-\rmi q^1 x^1}\,\rme^{-\frac{1}{2}eB\bigl[
  (x^1-\frac{p^2}{eB})^2+(x^1-\frac{p^{\prime 2}}{eB})^2\bigr]}
 \notag\\
 &= \exp\biggl\{ -\frac{1}{4eB}\Bigl[ (q^1)^2+(q^2)^2
  +4\rmi q^1 p^2 -2\rmi q^1 q^2 \Bigr] \biggr\}
\end{align}
for the $x^1$-integration.  Together with the contribution from the
$y^1$-integration we find,
\begin{equation}
 \exp\biggl\{ -\frac{1}{4eB}\Bigl[ (q^1)^2+(k^1)^2
  +2(q^2)^2 + 2\rmi(q^1+k^1)(2p^2-q^2) \Bigr]\biggr\} .
\end{equation}
Then, since the fermion propagator does not depend on $p^2$, it is
straightforward to perform the $p^2$-integration and we get
$\int\frac{\rmd p^2}{2\pi}\,\exp[-\frac{\rmi}{eB}(q^1+k^1)p^2]
=2\pi\delta(q^1+k^1)\cdot\frac{eB}{2\pi}$.  This is how we can obtain
the momentum conservation and the overall factor $eB/2\pi$ that is
exactly the Landau degeneracy factor.

After all, in the dimensional regularization, the polarization tensor
can be expressed as
\begin{equation}
 \Pi^{\mu\nu}(k,q) = (2\pi)^4\delta^{(4)}(k\!+\!q)\,
  \rme^{-\frac{1}{2eB}[(q^1)^2+(q^2)^2]}\, \Pi^{\mu\nu}(q)
\end{equation}
with
\begin{equation}
 \Pi^{\mu\nu}(q) = \biggl( g^{\bmu\bnu}-\frac{q^{\bmu}q^{\bnu}}{\bar{q}^2}
 \biggr)\frac{e^2|eB|}{2\pi^2} I(\bar{q}/m) ,
\label{eq:Pi}
\end{equation}
where $\Pi^{1\nu}=\Pi^{2\nu}=\Pi^{\mu1}=\Pi^{\mu2}=0$, which is
concluded from $P_0\gamma^1 P_0=P_0 \gamma^2 P_0=0$.  In the final
expression we have defined a function $I(\bar{q}/m)$ by
\begin{equation}
 I(x) = \int_0^1 \rmd y\, \frac{y(1-y)}{y(1-y) -x^{-2}}
  = 1-\frac{4\sin^{-1}(x/2)}{x\sqrt{4-x^2}} ,
\label{eq:def_I}
\end{equation}
for $x<2$, namely, $|\bar{q}|<2m$.  This should be expressed as
\begin{equation}
 I(x) = 1-\frac{2}{x\sqrt{x^2-4}} \biggl(\ln\biggl|
  \frac{x-\sqrt{x^2-4}}{x+\sqrt{x^2-4}} \biggr| + \rmi\pi\biggr),
\end{equation}
for $x>2$, namely, $|\bar{q}|>2m$ with $q^0>0$ assumed.  [Because we
  are considering the retarded self-energy, the sign of the imaginary
  part changes if $q^0<0$.]  An imaginary part appears in this case
because the gauge particle can decay into a fermionic particle and an
anti-particle above the mass threshold.


\begin{figure}
 \includegraphics[width=\columnwidth]{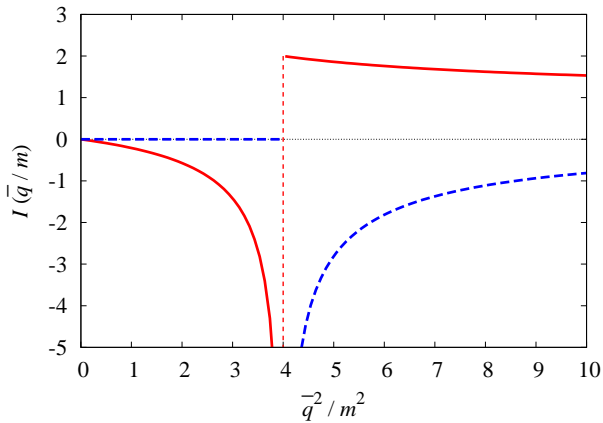}
 \caption{$I(\bar{q}/m)$ as a function of $\bar{q}^2/m^2$.  The solid
   and the dashed lines represent the real-part and the
   imaginary-part, respectively.  The mass threshold for
   particle--anti-particle decay is located at $\bar{q}^2/m^2=4$.}
 \label{fig:func}
\end{figure}


We make a plot in Fig.~\ref{fig:func} to show the behavior of the
function $I(\bar{q}/m)$.  We can clearly notice that the mass
threshold for particle--anti-particle decay is located at
$|\bar{q}|=2m$ where the real-part diverges if it is approached from
the smaller-$\bar{q}$ side, while the imaginary-part diverges if
approached from the larger-$\bar{q}$ side.  In the polarization tensor
in (3+1) dimensions there would be no such singularity at the mass
threshold because of the phase-space factor that compensates for the
divergence.

In the definition~\eqref{eq:def_I} we can readily take the limit of
$m\to0$, i.e.\ $x\to\infty$ to obtain $I(\infty)=1$.  Therefore, in
this limit, the polarization simply becomes
$\Pi^{\mu\nu}(q)=(g^{\bmu\bnu}-q^{\bmu}q^{\bnu}/\bar{q}^2)e^2|eB|/2\pi^2$.
This result has an interpretation as the (1+1)-dimensional self-energy
$(g^{\bmu\bnu}-q^{\bmu}q^{\bnu}/\bar{q}^2)m_\gamma^2$ multiplied by
the Landau-level degeneracy $|eB|/2\pi$, where $m_\gamma^2=e^2/\pi$ is
the ``photon mass'' in the massless Schwinger model
\cite{Schwinger:1962tp}.  Since the massless Schwinger model is
exactly solvable and the dimensional reduction is complete as we
checked above, this result at $m=0$ is presumably exact.

Here we shall make a rather technical remark about the momentum
integration.  The na\"{i}ve momentum integration for the one-loop
polarization in (1+1) dimensions is UV finite.  Nevertheless, the
final results could depend on a choice of the regularization
procedure.  To see this, let us take an example of
$\Pi^{00}(q^0=0,|\bq|\to0)$ in the $m=0$ limit.
From Eq.~\eqref{eq:Pi} we already know that the correct answer should
be $e^2|eB|/2\pi^2$ in this limit.  If we perform the
$p^0$-integration first \cite{Fukushima:2009ft}, however, we can
easily prove,
\begin{equation}
 \lim_{q\to0}\int\frac{\rmd p^0}{2\pi}\frac{2(p^0)^2-\bar{p}\cdot
  (\bar{p}-\bar{q})}{\bar{p}^2(\bar{p}-\bar{q})^2} = 0 .
\label{eq:integ}
\end{equation}
Then, one may conclude that $\Pi^{00}(q^0=0,|\bar q|\to0)=0$ which is
obviously different from the correct answer in the dimensional
regularization.  One can argue the computation of
$\Pi^{33}(q^0=0,|\bar q|\to0)$ in the same way to find a discrepancy
again.  Such subtlety originates from the fact that the limit of
$|\bar q|\to0$ is not commutative to the limit of $m^2\to0$ as is
noticeable from Fig.~\ref{fig:func}.  In other words the physical
properties in the region with $|\bar q|<2m$ are not accessible at all
by the knowledge from the massless Schwinger model.  We pay our
attention to this subtlety since it is common to process the
$p^0$-integration first in finite-$T$ calculations.  The gauge
invariant results~\eqref{eq:Pi} implies that the electric-current
susceptibility at vanishing momenta is zero, while the na\"{i}ve
estimates are not \cite{Fukushima:2009ft}.  Moreover, interestingly, a
statement similar to Eq.~\eqref{eq:integ} is correct for the matter
part involving the Fermi-Dirac distribution function at finite $T$.
We can explicitly make sure that the matter part with $m=0$, $q^0=0$,
and $|\bq|\to0$ is vanishing just like Eq.~\eqref{eq:integ}.  The
finite-$T$ extension will be reported elsewhere.
\vspace{3mm}


Now we can write the inverse propagator of gauge particles as
\begin{equation}
 \begin{split}
 -\rmi D_R^{-1\mu\nu}(q) &= g^{\mu\nu}q^2 - q^\mu q^\nu -
 \Pi^{\mu\nu}(q) \\ &\qquad\qquad +\text{(gauge fixing terms)} .
 \end{split}
\end{equation}
We note that we neglect the exponential factor
$\rme^{-\frac{1}{2eB}[(q^1)^2+(q^2)^2]}$ assuming that the transverse
external momenta are sufficiently smaller than $eB$, for which the LLL
approximation can be justified.  Let us clarify what collective
excitations are contained in the above propagator.  For this purpose
we should fix the gauge first and locate the pole(s) of
$D_R(\omega_q,\bq)$ as a function of $\omega_q$ for a given $\bq$.
Instead we can locate the pole position by solving the equation,
$\det[D_R^{-1}(\omega_q,\bq)]=0$.

Here we adopt the temporal axial gauge $A^0=0$ so that the physical
degrees of freedom is intuitively manifest.  Then, $D_R^{-1}$ is a
$3\times3$ matrix in the Lorentz index and we can show that
\begin{align}
 \det[D_R^{-1}]
 &= \det\biggl[ -q^2 \delta^{ij} - q^i q^j + \delta^{i3}\delta^{j3}
  \frac{(q^0)^2}{\bar{q}^2}\calM^2(q) \biggr] \notag\\
 &= -(q^0)^2 q^2 \bigl[ q^2 - \calM^2(q) \bigr] ,
\end{align}
where $\calM^2=(e^2|eB|/2\pi^2) I(\bar{q}/m)$.  The first factor
$(q^0)^2$ is from the longitudinal gauge field and this unphysical
contribution is exactly cancelled by the Faddeev-Popov determinant
associated with the $A^0=0$ gauge.  The next factor represents one of
two transverse gauge fields whose dispersion relation is intact.  The
final factor is the most interesting;  one of two transverse gauge
fields is influenced by the screening effect by $\calM^2(q)$.


\begin{figure}
 \includegraphics[width=\columnwidth]{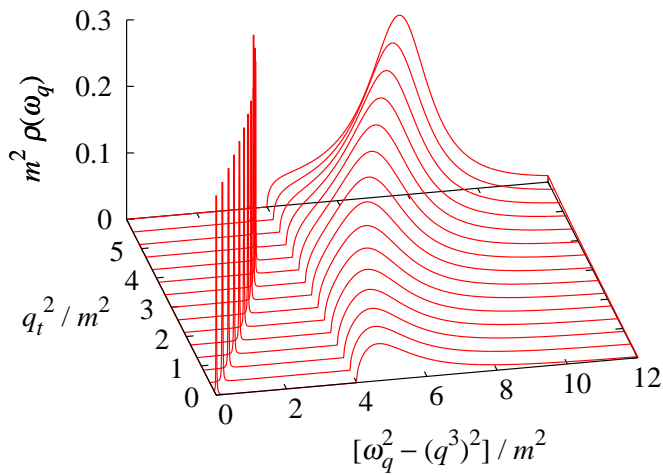}
 \caption{Spectral function which is made dimensionless by $m^2$.  We
   choose $e^2|eB|/(2\pi m^2)=1$.}
 \label{fig:rho}
\end{figure}


Solving $q^2-\calM^2(q)=0$, we can figure out the energy dispersion
relation $\omega_q(\bq)$ of the collective excitation in the same way
as reading the plasmon dispersion at finite temperature.  For
intuitive understanding we do not show the pole position but discuss
the spectral function which contains the information of both the
energy dispersion relation and the decay width.  Instead of defining
the spectral function from $D_R(q)$ directly, we define
$\rho(\omega_q)$ for the transverse mode affected by $\calM^2(q)$ as
if it were a scalar field, that is,
\begin{equation}
 \rho(\omega_q) = -\frac{1}{\pi} \frac{\text{Im}\calM^2(q)}
  {[q^2-\text{Re}\calM^2(q)]^2 + [\text{Im}\calM^2(q)]^2} .
\end{equation}
Figure~\ref{fig:rho} is the numerical result when we set
$e^2|eB|/(2\pi m^2)=1$.  Actually changing $e^2|eB|/2\pi m^2$ makes no
qualitative difference.  We can see that there are two separate
branches below and above the mass threshold.  In what follows we
discuss them in order.

Let us look at the branch above the threshold, i.e.\
$|\bar q|=\sqrt{\omega_q^2-(q^3)^2}>2m$.  In this case the collective
excitation (photon) is attenuated because it can decay into a particle
and an anti-particle.  The width is divergently large near the
threshold, which is already clear from the dashed line in
Fig.~\ref{fig:func}.  As the transverse momentum
$q_t^2=(q^1)^2+(q^2)^2$ increases the width becomes narrower and
narrower.  In the limit of $q_t^2\gg m^2$, eventually, the dispersion
is reduced to that of massive photon with its mass given by
$e^2|eB|/2\pi^2$.  This asymptotic situation is, as already mentioned,
the limit toward the massless Schwinger model.

We can notice that the spectral height grows larger as $q_t^2$
increases.  The spectral sum rule should be preserved with the
contribution from another branch below the mass threshold.  As noticed
from Fig.~\ref{fig:rho} this collective mode below the threshold lies
in the space-like region ($\omega_q^2<|\bq|^2$) without attenuation,
which looks like a sound mode.  Such an observation of the sound-like
mode has been also reported recently in the strong-coupling regime in
the AdS/CFT analysis \cite{Kharzeev:2010gd}.  Since we are working
diagramatically and the physical contents are transparent, we can give
a clear interpretation for this collective excitation in our
perturbative approach;  we propose that this should be an analogue of
the \textit{zero sound} in the Fermi liquid theory \cite{abrikosov}.

Because of the dimensional reduction, on the one hand, physical
properties of the system in a strong magnetic field looks like (1+1)
dimensional.  At finite density, on the other hand, physical
properties are dominated by the interactions near the Fermi surface,
so that the dynamics is effectively (1+1) dimensional too.  [Since the
  momentum in a certain direction is as large as the chemical
  potential, the momenta in other directions perpendicular to the
  motion are negligible and the system is locally (1+1) dimensional if
  the Fermi momentum is large enough.]  In condensed matter physics
the zero sound appears in Fermi liquid even at zero temperature as a
result of the particle--hole excitation in pseudo (1+1) dimensions
where the polarization has a logarithmic singularity similar to our
Fig.~\ref{fig:func}.  In the present case, the magnetic field instead
of the Fermi surface causes the pseudo (1+1) dimensionality, and the
particle--anti-particle excitation instead of the particle--hole
excitation constitute sound-like collectivity in the vector channel.

Usually the zero sound is a longitudinal sound mode associated with
the density fluctuation which appears only when the Landau parameter
$F_0$ in the fermionic interaction is repulsive.  In relativistic
matter the zero sound emerges as a gapless branch of the vector meson
\cite{Chin:1977iz}.  It is remarkable that in our case the
zero-sound-like mode lies in the transverse degrees of freedom, so
that it is indistinguishable from the massless photon even though the
microscopic content is totally different.

In this work we focus on the $B$-induced screening effects on massless
gauge particles only.  It would be very interesting to think about the
screening effects on massive vector mesons in the same way as in
Ref.~\cite{Chin:1977iz}.  We would then anticipate that a massless
collective mode should exist in the space-like region as discussed in
Ref.~\cite{Chin:1977iz}, which is the complete counterpart of the zero
sound, which we name the ``magnetic zero sound'' here.  Recently it
was pointed out that the $\rho$-meson superconductor could be possible
in the vacuum with a superstrong $B$ field \cite{Chernodub:2010qx}.
It is an important question how this electric superconductivity should
or should not be modified by the screening effects and particularly by
the existence of the magnetic zero sound.


In summary, in the LLL approximation, we found that one of transverse
gauge particles is screened by the magnetic field, which leads to two
branches of the collective excitations.  One has a modified dispersion
relation with a mass and a width whose behavior is reminiscent of the
plasmon at finite temperature.  The other looks like the zero sound at
finite density, which can be naturally understood from the pseudo
(1+1) dimensional nature under a strong magnetic field.  These modes
are relevant to the situation with $|eB|\gg m^2 \sim |\bq|^2$ and
convey detailed information of the birefringence in the vacuum (not
medium) with a magnetic field background.  Because a huge $|eB|$ is
expected in the heavy-ion collision, we may have a chance to probe
these collective modes indeed.  If we replace the overall coupling
constant $e^2$ in $\calM^2$ by $g^2$ for quark loops, it would enhance
the chance for observation and also we can discuss the physics
implication of our results to the QCD phenomena such as the chiral
magnetic effect.  Using the NJL-type models we can also discuss the
realization of the magnetic zero sound in the excitation spectra of
vector mesons.  These applications are interesting future problems.
\vspace{5mm}


The author thanks Wolfram~Weise for useful comments and his kind
hospitality at TUM where this work was complete.  He also thanks
Dmitri~Kharzeev and Harmen~Warringa for discussions that inspired him
toward this work.  He is grateful to Maxim~Chernodub for valuable
comments.  He was supported by Grant-in-Aid for Young Scientists B
(No.\ 20740134).


\end{document}